\documentclass[12pt,a4paper]{article}
\usepackage{t1enc}
\usepackage[latin1]{inputenc}
\usepackage[english]{babel}
\usepackage{graphicx}
\font\bBB=msbm10
\font\bBBB=msbm10 at 16pt

\def\p{\par}
\def\bBC{\mbox{\bBB C}}

\def\bBN{\mbox{\bBB N}}

\def\bBR{\mbox{\bBB R}}

\def\bBZ{\mbox{\bBB Z}}

\def\bBBZ{\mbox{\bf \bBBB Z}}
\begin{document}
\title {Modular Conjugation and the Implementation of Supersymmetry}
\author{
 Orlin Stoytchev\thanks{American University in Bulgaria, 2700 Blagoevgrad, Bulgaria\newline and Institute for Nuclear Research, 1784 Sofia, Bulgaria, \quad ostoytchev@aubg.bg} }
\date{}
\maketitle
\abstract{
Any $\bBZ_2$-graded $C^*$-dynamical system with a self-adjoint graded-KMS functional on it can be represented (canonically) as a $\bBZ_2$-graded algebra of bounded operators on a $\bBZ_2$-graded Hilbert space, so that the grading of the latter is compatible with the functional. The modular conjugation operator plays a crucial role in this reconstruction. The results are generalized to the case of an unbounded graded-KMS functional having as dense domain the union of a net of $C^*$-subalgebras. It is shown that the modulus of such an unbounded graded-KMS functional is KMS.
}
\section{Introduction}
\par\smallskip
The notion of a super-KMS  functional \cite{Jaffe} and  more generally a graded-KMS functional \cite{Kastler, Longo}
have been introduced mainly in connection with the observation of their importance to
cyclic cohomology theory \cite{Jaffe2,Connes, Kastler}.  As advocated in \cite{Kastler2}, the super-KMS
functionals seem to be the appropriate substitute for elliptic operators when
one passes from finite-dimensional to infinite-dimensional and (or) from commutative to
noncommutative geometry, in the sense of \cite{BConnes}.\p
A prototype for a graded-KMS functional on the algebra of bounded operators on a $\bBZ_2$-graded Hilbert space is given by a regularized supertrace
$$\omega(\ \cdot\ )=\hbox{str}(\ \cdot\,\rho)$$
with $\ \rho\ $--- an even positive trace-class operator.  Such functionals appear in finite-volume supersymmetric quantum field theories in thermal background and have been used to define the Witten index \cite{Witten, Longo2}.\p
In an abstract $C^*$-dynamical setting the graded-KMS condition is a natural supersymmetric generalization of the KMS (Kubo, Martin, Schwinger) condition. 
It is defined with respect to a grading $ \gamma $ of
the $C^*$-algebra $ {\cal A} $ and a continuous one-parameter *-automorphism group
$ \alpha_t $, commuting with $ \gamma $.  Namely $ \omega $ is a {\it graded-KMS functional on $ {\cal A} $} if
it satisfies
\begin{equation}
\omega(ab)=\omega(b^{\gamma}\,\alpha_i(a)) \label{skms}
\end{equation}
for any analytic with respect to $ \alpha_t $ element $\ a\in{\cal A}\ $ and any $b\in{\cal A} $. The grading $ \gamma $ acts as $ b\rightarrow b^{\gamma}:=b_+-b_- $,
where $ b_{\pm} $ are the even and odd parts of $ b $ respectively and
$ \alpha_i(a) $ is the value of $\ \alpha_z(a)\ $ at $\ z=\sqrt{-1}\ $.\p
As shown in \cite{me}, condition (\ref{skms}) arises naturally in the case when $\omega$ is a
faithful normal (nonpositive) self-adjoint functional on a von
Neumann algebra $ {\cal A} $. More precisely there exist a canonical $\sigma$-weakly
continuous one-parameter group $\ \alpha_t\ $ (the ``modular group'') and a
canonical
$\bBZ_2$ grading $\ \gamma\ $ on $ {\cal A} $, commuting with the automorphism group,
and $ \omega $ is graded-KMS with respect to them. Furthermore, in complete
analogy with the standard Tomita-Takesaki theory, (where $ \omega $ is assumed
positive instead of just self-adjoint,) the canonical automorphism group and
grading are the unique ones (with certain properties) with respect to which $ \omega $ is graded-KMS.\p
A more general notion was studied in \cite{Longo} --- that of a twisted-KMS functional. The defining relation (\ref{skms}) is the same but $\gamma$ is now an arbitrary *-automorphism, not necessarily an involution. It was shown in \cite{Longo} that some of the results proven for von Neumann algebras in \cite{me} remain valid in the more general $C^*$ setting and when ``graded'' is replaced by ``twisted''. In particular, if $\omega$ is twisted-KMS, its modulus, which is a positive functional, is KMS (when normalized to one).\p
The regularized supertrace, as a prototype for a graded-KMS functional, has two additional properties. It is self-adjoint, i.e., it is real on self-adjoint elements of  $ {\cal A} $. In addition, the grading of the Hilbert space ${\cal H}={\cal H}_+ \oplus {\cal H}_-$ is {\it compatible} with the functional $\omega$. By the latter we mean that for any projection $e\in {\cal A}''$ (the weak closure or double commutant of $ {\cal A} $), $\omega(e)\geq 0$  if $\hbox{ran}\,e\subseteq {\cal H}_+$  and $\omega(e)\leq 0$  if $\hbox{ran}\,e\subseteq {\cal H}_-$.\p
It is interesting to find out the extent to which the regularized supertrace is a generic example of a graded-KMS functional. The present paper tries to answer this. In other words, we study the following question: Given an abstract $\bBZ_2$-graded $C^*$-algebra 
$ {\cal A} $ with an action of a continuous one-parameter *-automorphism group $ \alpha_t $, preserving the grading and a self-adjoint graded-KMS functional $\omega$, can we represent $ {\cal A} $ as a $\bBZ_2$-graded algebra of operators on a $\bBZ_2$-graded Hilbert space ${\cal H}$, so that the grading of ${\cal H}$ is compatible with $\omega$ and the grading of 
$ {\cal A} $ is induced from that of $ {\cal H} $? Further, is $\omega$ a regularized supertrace in that representation?
Another way is to say that  we would like to reconstruct an abstract $\bBZ_2$-graded $C^*$-dynamical system with a graded-KMS functional on it and we notice that the grading of the Hilbert space is encoded in the functional, not in the grading of the algebra and in any ``good'' representation the grading of the algebra should be the one induced from that of the Hilbert space. 
\p
We believe that the reconstruction result of this paper has relevance to the
transition to infinite volume limit in supersymmetric finite volume quantum
field theories.
The principal difficulties that one has to overcome are two --- reconstruction of
the Hilbert space with a nonpositive functional $\omega$ and finding a $\bBZ_2$ graded
representation with respect to a  $\bBZ_2$ grading of $ {\cal H} $, compatible with $\omega$. 
The first issue is addressed easily by replacing $\omega$ with the (canonically defined) positive functional $|\omega|$, called the modulus of $\omega$. The GNS construction with $|\omega|$ and some of its properties will be discussed in the next section.
It turns out that this standard GNS representation $\pi$ has a flaw. It is not a $\bBZ_2$ graded
representation with respect to a  $\bBZ_2$ grading of $ {\cal H} $, compatible with $\omega$. In fact the whole algebra is mapped 
onto even operators on $ {\cal H} $. \p
Somewhat miraculously, there exists an antiunitary operator $J$ on the GNS Hilbert space with just the right properties, so that if we form a unitary $U:=K\,J$, where $K$ is complex conjugation in some basis, then $\pi':=U\,\pi\,U^*$ is a proper $\bBZ_2$ graded representation. Even though $J$ is defined directly in our work, without any reference to Tomita-Takesaki's theory, it coincides with the modular conjugation operator in the
case, when $\omega $ is faithful. These developments are the subject of Section 3.\p
Section 4 generalizes the reconstruction results to the case when $\omega$ is unbounded graded-KMS functional with a dense domain in ${\cal A}$. The need to consider this more general case stems from the results of, and has been suggested by, Buchholz, Longo and others 
\cite{Longo, Buch}, who showed that the graded-KMS condition  for bounded functionals is incompatible with the existence of an automorphism group $\alpha_t$ acting in an asymptotically abelian way, which is the typical situation of a local quantum theory in thermal background in infinite volume.\par
Motivated by the physical context, we take the domain of $\omega$ to be 
${\cal A}_{\hbox{\tiny loc}}:=\bigcup_{\cal O} {\cal A(O)}$, the algebra of local ``observables'',
and the algebra ${\cal A}$ to be the completion of ${\cal A}_{\hbox{\tiny loc}}$. Here ${\cal A(O)}$ is an increasing net of $C^*$ subalgebras. We show that  the modulus $|\omega|$ of an unbounded graded-KMS functional $\omega$ is a positive unbounded KMS functional, thus generalizing results of \cite{me, Longo}.
\section{The GNS construction}
In this section we put together mostly known facts and prepare the ground
for our main result in Section 3.
Let ${\cal A}$ be a unital $C^*$-algebra and $\omega$ --- a self-adjoint continuous linear
functional on
it. We list some definitions and facts we shall need.\p
Two positive functionals $\varphi$ and $ \psi $ are called orthogonal (denoted
$ \varphi\,\perp\,\psi $) if the following equality holds:
$$
\|\varphi-\psi\|\,=\,\|\varphi\|\,+\,\|\psi\|\,.
$$
Every self-adjoint functional $ \omega $ on a $C^*$-algebra has a unique
decomposition into two orthogonal positive functionals $\ \omega_{\pm}\ $, called
{\it Jordan decomposition} (\cite{Ped}, Sec. 3.2):
\begin{equation}
\omega\,=\,\omega_+\,-\,\omega_-\ ,\quad\omega_+\perp\omega_-\,.\label{Jord}
\end{equation}
The Jordan decomposition of a self-adjoint functional $ \omega $ is preserved by
any  *-automorphism that leaves $ \omega $ invariant, i.e.
$$
\omega\circ\alpha\,=\,\omega\quad\Longleftrightarrow\quad\omega_{\pm}\circ\alpha\,=\,\omega_{\pm}\,.
$$
This follows easily from the fact that *-automorphisms preserve positivity and mutual orthogonality of functionals and from the uniqueness of the Jordan decomposition.\p
One can associate a (unique) positive functional $|\omega |$ to the self-adjoint $\omega$:
\begin{equation}
|\omega |:=\omega_+ + \omega_-\ .\label{mod1}
\end{equation}
The positive functional $|\omega |$ is called {\it the modulus} of $\omega$  and can be defined in fact for an arbitrary $\omega$ as the unique positive functional, satisfying (see \cite{Dix}, Sec. 12.2.9.):
\begin{equation}
\|\,|\omega |\,\|=\|\omega\|\ ,\quad |\omega(a)|\leq \|\omega\|\,|\omega |(a^*a)\ ,\ \ a\in{\cal A}\ .\label{mod2}
\end{equation}
An easy exercise shows that the functional, defined by (\ref{mod1}) satisfies the conditions in (\ref{mod2}).\p
Consider now the Gelfand - Naimark - Segal (GNS) construction $(\pi, {\cal H}, \Omega )$ associated with the positive functional 
$|\omega|$. Any element $a\in {\cal A}$ is represented by a bounded operator $\pi(a)$ on the Hilbert space ${\cal H}$ and we have $|\omega|(a)=(\Omega, \pi(a)\,\Omega)$. The algebra $\pi({\cal A})$ is a *-homomorphic image of ${\cal A}$ which is in general not isomorpic to ${\cal A}$, since 
$|\omega|$ is not necessarily faithful.  \p
The positive functionals $\omega_+$ and $\omega_-$ are dominated by $|\omega |$ and therefore (see \cite{Ped}, Sec. 3.3.) there are unique elements $p_{\pm }\in \pi({\cal A})'$ with $0\leq p_{\pm}\leq 1$, such that for all $a\in {\cal A}$
\begin{equation}
\omega_{\pm }(a)=(\Omega, \pi(a)\,p_{\pm }\,\Omega)\ .\label{omegapm}
\end{equation}
Since $\omega_1 +\omega_2 = |\omega|$, we have $p_+ + p_- = 1$ and since $\omega_+\perp\omega_-$ the elements
$p_{\pm}$ are actually mutually orthogonal projections.\p
The latter statement is obvious in the commutative case, when the Jordan decomposition of a self-adjoint functional reduces to the usual Jordan decomposition of a signed measure and the elements $p_{\pm}$ are the characteristic functions of the sets on which the measure is positive and negative, respectively. In the general case one way to prove that $p_{\pm}$ are projections is as follows:\p
Consider the spectrum $\sigma$ of the operator $p_+$. We have $\sigma\subset [0,1]$. Suppose that we can find  
$\delta > 0$ so that the spectral projection $p_0$, corresponding to the interval $[\delta, 1/2]$ is not zero. Note that since 
$p_-=1-p_+$, we have 
$p_-\,p_0\geq p_+\,p_0$.
It is known (\cite{Ped}, Sec. 3.2.) that two positive functionals
$\omega_+$ and $\omega_-$ are orthogonal if and only if for every $\epsilon >0$ there is a positive element $z$ in the unit ball of ${\cal A}$, such that  $\omega_+(1-z)<\epsilon $ and $\omega_-(z)<\epsilon $. We have 
$$
(\Omega, (1-z)\,p_+\,p_o\,\Omega)\leq (\Omega, (1-z)\,p_+\,\Omega)=\omega_+(1-z)<\epsilon 
$$
and therefore

$$
(\Omega, z\,p_+\,p_0\,\Omega)>(\Omega,p_+\,p_0\,\Omega)-\epsilon \ .
$$
But then 
$$
\omega_-(z)\geq (\Omega,z\,p_-\,p_0\,\Omega)\geq (\Omega,z\,p_+\,p_0\,\Omega)>(\Omega,p_+\,p_0\,\Omega)-\epsilon \ ,
$$
which contradicts the statement that $\omega_-(z)$ can be made arbitrarily small. The contradiction shows that $\sigma\cap [\delta, 1/2]=\emptyset$. Similarly, by interchanging the roles of $\omega_+$ and $\omega_-$ we show that $\sigma\cap 
[1/2, 1-\delta ]=\emptyset$. Therefore $\sigma=\{0,1\}$ and $p_+$ and $p_-$ are projections.\p\smallskip
There is an obvious $\bBZ_2$-grading (orthogonal decomposition) of the Hilbert space --- ${\cal H}={\cal H}_+ \oplus {\cal H}_-$ with ${\cal H}_{\pm}:=p_{\pm}\,{\cal H}$. The functional $\omega$ is expressed as a ``graded vacuum expectation value'':
\begin{equation}
\omega(a)=(\Omega, \pi(a)\,(p_+-p_-)\,\Omega)=(\Omega, \pi(a)\,\Gamma\,\Omega)\ ,\label{Gamma}
\end{equation}
where the grading operator $\Gamma$ has the bock-diagonal form
$$\Gamma\,=\left(\begin{array}{c c}
I&0 \\
0&-I \end{array}\right)\ .
$$
This grading of ${\cal H}$ is obviously compatible with $\omega$. Since every $\pi(a)$ commutes with $\Gamma$, the representation is reducible and all $\pi(a)$ have block-diagonal form, i.e., they are all even operators relative to this grading.
Thus the GNS representation $(\pi, {\cal H}, \Omega )$ is not a $\bBZ_2$-graded representation of ${\cal A}$\p
{\bf Note} One may be misled to think that equation (\ref{Gamma}), relating $\omega$ to $|\omega|$ represents the polar decomposition \cite{Sakai} of a self-adjoint functional, but this is wrong. The  polar decomposition (of a normal functional on a von Neumann algebra) relates a functional  to its modulus via an element of the algebra, while the operator $\Gamma$ is in the commutant of 
$\pi({\cal A})$. In the $C^*$ setting the polar decomposition of $\omega$ involves an element $g$ of the weak closure of  $\pi({\cal A})$, ( i.e. its double commutant) as discussed in \cite{Longo} and further in our paper. This element $g$ does not commute with $\pi({\cal A})$ unless ${\cal A}$ is trivially graded. In fact $g$ implements the grading automorphism.\p
With a slight abuse of notations we will write $|\omega|$, $\omega$ and $\omega_{\pm}$ for the extensions of the respective functionals, using their GNS representations, to the von Neumann algebra ${\cal B}:=\pi({\cal A})''$ (the weak closure of
$\pi({\cal A})$). Thus we have
\begin{equation}
|\omega|(a):=(\Omega, a\,\Omega)\ ,\quad \omega(a):=(\Omega, a\,\Gamma\,\Omega)\ ,\quad
\omega_{\pm}(a):=(\Omega, a\,p_{\pm}\,\Omega)\ ,\quad a\in{\cal B}\ .\label{omegas}
\end{equation}
It is trivial that all four functionals are normal (or $\sigma$-weakly continuous), that $\omega$ is self-adjoint with  $|\omega|$ being
its modulus and $\omega_{\pm}$ giving its Jordan decomposition.
In general, the Jordan
decomposition $ \omega=\omega_+-\omega_-\ $ of a normal self-adjoint functional  has the following additional properties
\cite{Dix2}:\\
(i) $\omega_{\pm} $ are normal positive functionals with mutually singular
supports, i.e.,
any $ a\in{\cal B},\ a\ge 0 $ can be represented (nonuniquely in general) as a sum
$\ a=a_++a_-\ $ so that $\ \omega_-(a_+)=\omega_+(a_-)=0\ $.\\
(ii) There exist projections $\ \chi_{\pm}\in{\cal B}\ $ (not necessarily unique) onto
the supports of
$\ \omega_{\pm}\ $ with the following properties:
\begin{eqnarray}
&\chi_+\chi_-=\chi_-\chi_+=0\,,\\
&\omega_+(a)=\omega(a\chi_+)\,,\ \ \quad a\in{\cal B}\,,\label{chi1}\\ 
&\omega_-(a)=-\omega(a\chi_-)\,,\ \ \quad a\in{\cal B}\,.\label{chi2}
\end{eqnarray}
Using these projections and writing $g:=\chi_+-\chi_-$ one can link the modulus $ |\omega| $ to $ \omega $:
\begin{equation}
|\omega|\,=\,\omega_++\omega_-\,=\,g\circ \omega\,. \label{Sakai1}
\end{equation}
where $\ g\circ\omega\ $ is just a notation for the functional defined
by 
$ g\circ\omega(a):=\omega(a\,g) $. We also have quite easily:
\begin{equation}
\omega=g\circ |\omega|\,.\label{Sakai2}
\end{equation}
Formulae (\ref{Sakai1}) and (\ref{Sakai2}) are indeed a special case of the polar decomposition \cite{Sakai}
of a normal linear functional.\p
The projections $ \chi_{\pm} $ are not unique if and only if $ |\omega| $ is not faithful.
In the case we have at hand $|\omega|=(\Omega, \,\cdot \  \Omega)$ is faithful on $\pi({\cal A})$ (even though $|\omega|$ may not be faithful as a functional on ${\cal A}$). By construction $\Omega$ is cyclic for $\pi({\cal A})$ and the existence (see next section) of the antiunitary operator $J$ (modular conjugation) a posteriori shows that $\Omega$ is also cyclic for $J\pi({\cal A})J=\pi({\cal A})'$. This implies that $\Omega$ is separating for $\pi({\cal A})''={\cal B}$ which is the same as saying that $|\omega|$ is faithful as a functional on ${\cal B}$.
Thus $ \chi_{\pm} $ are unique and invariant under any
*-automorphism $ \alpha $ which leaves $ \omega_{\pm} $ invariant. In addition
$\chi_++\chi_-=1$ and $g^2=1$. (The considerations in the next section remain valid in the case when $|\omega|$ is not faithful, as long as we choose $\chi_{\pm} $ to be the unique  minimal projections, satisfying (\ref{chi1}) and (\ref{chi2}). Of course in this case $\chi_+ +\chi_-\ne 1$ and $g^2\ne 1$.)
\section{The ${\bf \bBBZ_2}$-graded representation}
It turns out that the graded-KMS
property of the functional $ \omega $ gives a natural way to define a conjugate
representation $ \pi' $, which unlike $\pi$ respects the grading of $ {\cal A} $.\p
First note that any *-automorphism of $ {\cal A} $, preserving $|\omega|$ can be implemented in the GNS Hilbert space by the adjoint action of a unitary operator. In this way the action of this *-automorphism can be extended from $ {\cal A} $ to 
${\cal B}=\pi({\cal A})'' $. In particular, ${\cal B}$ becomes a $\bBZ_2$-graded von Neumann algebra with a strongly continuous *-automorphism group $\alpha_t$, preserving the grading.\p
Using approximation arguments, it is shown in \cite{Longo} (Lemma 1) that the extension of $\omega$ to ${\cal B}$ (which we denote by the same letter) is a graded-KMS functional.\p
We state now some important results for graded-KMS normal self-adjoint
functionals. For the proofs see \cite{me} or \cite{Longo}.\p
\smallskip\noindent 
{\it Let $\ \omega\ $ be a graded-KMS normal self-adjoint functional on the $\bBZ_2$
graded von Neumann algebra $ {\cal B}={\cal B}_+\oplus{\cal B}_- $
 and $ \omega_{\pm}$, $\ |\omega|\ $ are defined as in Section 2. Then the following identities hold:
\begin{eqnarray}
\omega_-(b\chi_+a)=\omega_+(b\chi_-a)=0,\qquad a\in {\cal B}_+,\ b\in{\cal B}, \label{id1}\\
\omega_+(b\chi_+a)=\omega_-(b\chi_-a)=0,\qquad a\in {\cal B}_-, \ b\in{\cal B}.\label{id2}
\end{eqnarray}
The functional $\ |\omega|\ $ is a KMS functional. If $ a\in{\cal B}_+ $ is in
the left
kernel of $ \omega_+ $, then $ a^* $ is in the left kernel of $ \omega_+ $ as well.
The same is true for $ \omega_- $. If $ a\in{\cal B}_- $ is in the left kernel of
$ \omega_+ $, then $ a^* $ is in the left kernel of $ \omega_- $ and vice
versa.}\p   \smallskip\noindent
A functional $ \phi$ is called even with respect to the grading $\gamma$ if
$\phi(a^{\gamma})=\phi(a),\ \  a\in {\cal B} $. It is easy to see that the
graded-KMS property, applied to $\omega(1\,a)$, implies that $ \omega $ is even and therefore, since the
action of $ \gamma$ is a *-automorphism, $ \omega_{\pm} $ and $ |\omega| $ have to
be even too.\p
A simple calculation shows that $p_{\pm}\,a\,\Omega=a\,\chi_{\pm}\,\Omega$ (but $\not= \chi_{\pm}\,a\,\Omega$ ) for any $a\in {\cal B}$ and therefore
$$
{\cal H}_{\pm}\equiv p_{\pm}\,{\cal H}=(\pi({\cal A})\,\chi_{\pm}\,\Omega)^-\ ,
$$
(where $(\cdot )^-$ signifies completion).
We shall need a further decomposition  of each $ {\cal H}_{\pm} $  into orthogonal direct sums.
Define the subspaces $\ {\cal H}^0_{\pm}:=(\pi({\cal A_+})\,\chi_{\pm}\,\Omega)^-$ and $\ {\cal H}^1_{\pm}:=(\pi({\cal A_-})\,\chi_{\pm}\,\Omega)^-$.
 The subspaces $\ {\cal H}^{0,1}_+\subset {\cal H}_+\ $ are mutually
orthogonal and so are $\ {\cal H}^{0,1}_- \subset {\cal H}_-\ $, i.e. $\ {\cal
H}^0_{\pm}
\perp {\cal H}^1_{\pm}\ $. We show that orthogonality holds for the dense
subspaces.
Take e.g., $ a\in {\cal A}_+ $ such that $\omega_-(a^*a)=0 $ and $ b\in {\cal A}_- $ such that
$\omega_-(b^*b)=0$. Then, because $ \omega_+ $ is even
we get:
$$
(\pi(a)\,\Omega,\pi(b)\,\Omega)=|\omega|(a^*b)=\omega_+(a^*b)=0\ .
$$
\p\bigskip
\noindent
{\bf Definition:}  Define an operator $J\ :{\cal
H}
\rightarrow {\cal H}\ $ by its action on a dense subspace:
$$
J\,\pi(a)\,\Omega:=\pi(\alpha_{\frac i2}(a^*))\,\Omega\ ,\quad a \hbox{\ \ analytic in } {\cal A}\ .
$$
\p
Those familiar with the Tomita--Takesaki theory will realize that the operator
$ J $  coincides with the {\it modular conjugation operator} in that theory.
Recall that  one defines an antilinear operator $ S $ as the closure of the operator
$ S_0\,\pi(a)\,\Omega:=\pi(a^*)\,\Omega $. Then the polar decomposition of $ S $ is
given by $ S=J\Delta^{\frac 12} $, where $ J $ is shown to be antiunitary
and $ \Delta $ is self-adjoint. (The {\it modular operator} $\Delta$ then is used to construct a one-parameter automorphism group 
--- the {\it modular group}
via the adjoint action of $\Delta^{i\,t}$ and it turns out that the functional $(\Omega,\,\cdot\,\Omega)$ is KMS with respect to that modular group.)\p
Our starting point is different however --- we have an automorphism
group to begin with and this allows us to define the modular conjugation $J$ in a very simple fashion.\par\bigskip\noindent
{\bf Proposition 1.} {\it $ J $ is an antiunitary operator, mapping $ {\cal
H}^1_+ $
onto $ {\cal H}^1_- $ and $ {\cal H}^1_- $ onto $ {\cal H}^1_+ $ and
leaving invariant separately $ {\cal H}^0_{\pm} $. Furthermore it is equal to
its inverse.}\par\smallskip\noindent
{\bf Proof:} First we show the last part:
$$
J^2\pi(a)\,\Omega=J\pi(\alpha_{\frac i2}(a^*))\,\Omega=\pi(\alpha_{\frac i2}(\alpha_{\frac i2}(a^*))^*)\,\Omega
=\pi(\alpha_{\frac i2}(\alpha_{-\frac i2}(a))\,\Omega)=\pi(a)\,\Omega.
$$
Next, we know from the results stated above that if $ \pi(a)\,\Omega\in {\cal
H}^1_+ $,
then $ \pi(a^*)\,\Omega\in{\cal H}^1_- $, while if $ \pi(b)\,\Omega\in{\cal H}^0_+ $ then.
$ \pi(b^*)\,\Omega\in{\cal H}^0_+ $ also. But the automorphisms $ \alpha_z $ leave
$ {\cal H}_{\pm} $
invariant (since $ \omega_+ $ and $ \omega_- $ are invariant separately), so
$ \pi(\alpha_{\frac i2}(a^*))\,\Omega\in{\cal H}^1_- $ and $ \pi(\alpha_{\frac i2}(b^*))\,\Omega
\in{\cal H}^0_+ $.
 $ J $ is obviously antilinear
since it
involves the antilinear operation $ a\rightarrow a^* $. Finally $ J $ is norm
preserving. Take any $ \pi(a)\,\Omega $. Then we calculate (remembering that $|\omega|$ is KMS):
\begin{eqnarray}
(J\,\pi(a)\,\Omega,J\,\pi(a)\,\Omega)=|\omega|((\alpha_{\frac i2}(a^*))^*\,\alpha_{\frac i2}
(a^*))=|\omega|(\alpha_{-\frac i2}(a)\,\alpha_{\frac i2}(a^*)) \nonumber\\
=|\omega|(a\,\alpha_i(a^*))=|\omega|(a^*\,a)=(\pi(a)\,\Omega,\pi(a)\,\Omega)\ .\nonumber
\end{eqnarray}
For the scalar product of two elements $ \pi(a)\,\Omega$ and $ \pi(b)\,\Omega$ we
obtain:
\begin{eqnarray}
(J\,\pi(a)\,\Omega,J\,\pi(b)\,\Omega)=|\omega|(a\,\alpha_i(b^*))=|\omega|(b^*\,a)\nonumber \\
=|\omega|((a^*\,b)^*)=\overline{|\omega|(a^*\,b)}=\overline{(\pi(a)\,\Omega,\pi(b)\,\Omega)}\ .\nonumber
\end{eqnarray}
which completes the proof.\p\bigskip\noindent
For every antiunitary map $ J $ there is a noncanonical unitary map $ U $ defined
as
$$U:=K\,J$$
where $ K $ is the operator of complex conjugation with respect to some chosen
orthonormal basis in the Hilbert space. As we shall see, different choices of bases lead to isomorphic representations. Since 
$K^2=I$ one can easily see that $U^*=JK$ and $U$ is unitary.\par
We observe the following properties of the different restrictions of $ U $ and
$ \pi(a) $:
\begin{equation}\begin{array}{ll}
U\, :\,{\cal H}^0_{\pm}\rightarrow{\cal H}^0_{\pm}\ ,\quad
&U\, :\,{\cal H}^1_{\pm}\rightarrow{\cal H}^1_{\mp}\ ,
\\
\pi(a)\,:\,{\cal H}^0_{\pm}\rightarrow{\cal H}^0_{\pm}\ ,\ \ 
&\pi(a)\,:\,{\cal H}^1_{\pm}\rightarrow {\cal H}^1_{\pm}\,\ \ a\in {\cal A}_+\ ,
\\
\pi(b)\,:\,{\cal H}^0_{\pm}\rightarrow{\cal H}^1_{\pm}\ ,\ \ 
&\pi(b)\,:\,{\cal H}^1_{\pm}\rightarrow{\cal H}^0_{\pm}\ ,\ \ b\in {\cal A}_-\ .
\end{array}\label{maps}
\end{equation}
{\bf Definition:} For every $ a\in{\cal A} $ define an operator
$ \pi'(a)\,:\,{\cal H}\rightarrow{\cal H} $ as follows:
$$
\pi'(a):=U\pi(a)U^*\ .
$$
We now prove the main result of this section.\par\bigskip\noindent
{\bf Proposition 2.} {\it The operators $ \pi'(a) $ are bounded for any $a\in {\cal A} $.
The map $ \pi':{\cal A}\rightarrow L({\cal H}_+\oplus{\cal H}_-) $ gives a
($\bBZ_2$ graded)
representation of the $\bBZ_2$ graded $C^*$-algebra $ {\cal A} $.}\p
\bigskip\noindent
Before proceeding with
the proof we would like to make the following comments.
A representation of a $\bBZ_2$ graded $C^*$-algebra $\ {\cal A}\ $ is by
definition a
*-algebra homomorphism
$ \pi'\,:\,{\cal A}_+\oplus{\cal A}_-\rightarrow
(L_+\oplus L_-)({\cal H}_+\oplus{\cal H}_-) $ ($L({\cal H})$ meaning all bounded operators on ${\cal H}$), which commutes with (preserves) the grading.
The grading $ L_+\oplus L_- $ is the natural one induced from the grading of
${\cal H}={\cal H}_+\oplus{\cal H}_-$.\newline
{\bf Proof:}\newline
(i) Algebra homomorphism:\newline
As $ \pi' $ is obviously linear, we only need to show that
$ \pi'(ab)=\pi'(a)\pi'(b)\,,\forall a,b\in{\cal A}\ $. This is obvious from
the definition of $ \pi' $ and the fact that $U$ is unitary. \newline
(ii)$ \pi' $ commutes with the grading:\newline
This is evident from the way $ \pi' $ was constructed. For $ a\in{\cal A}_+,\ 
\pi'(a)\,:\,{\cal H}_{\pm}\rightarrow{\cal H}_{\pm} $, i.e., $ \pi'(a)\in L_+({\cal H}_+\oplus{\cal H}_-) $,
and for $ b\in{\cal A}_-,\ \pi'(b)\,:\,{\cal H}_{\pm}\rightarrow{\cal H}_{\mp} $, i.e. $ \pi'(b)
\in L_-({\cal H}_+\oplus{\cal H}_-) $.\newline
(iii) A *-homomorphism:\newline
This is also immediate from the definition
$$
\pi'(a)^*=(U\,\pi(a)\,U^*)^*=U\,\pi(a)^*\,U^*
=U\,\pi(a^*)\,U^*=\pi'(a^*)\, 
$$
where we used the fact that the standard GNS representation $ \pi $ is a
*-homomorphism.\par\smallskip\noindent
{\bf Note} One can find mentioned in the literature (see, e.g., \cite{Haag}) the conjugate-linear representation $\pi_r:=J\pi J$ which is not a representation in the usual sense, i.e., it is not an algebra homomorphism. 
It is well known that conjugating an element $\pi(a)$ with the modular conjugation operator $J$ one gets an element in the commutant $\pi({\cal A})'$. Thus we have $\pi_r({\cal A})\equiv J\,\pi({\cal A})\,J\,\subset\pi({\cal A})'$. In general we do not have equality since $\pi({\cal A})$ is not a von Neumann algebra.
\par\bigskip\noindent
The next few statements have easy proofs which we omit.\p
In the representation $ \pi' $ the functional $ \omega $ is again expressed as a
graded vacuum expectation value.
$$\omega(a)=(\Omega,\Gamma\pi'(a)\Omega)\,.$$
\p 
The operator $g$ implementing the grading is mapped to $\Gamma$, when conjugated with $U$:
$$
U\,g\,U^*\equiv U\,(\chi_+ - \chi_-)\,U^*=\Gamma =\left(\begin{array}{cc}I&0\\ 0&-I\end{array}\right)
$$
\p
The vacuum $ \Omega $ is a cyclic vector for the representation $ \pi' $.\p
The next statement treats the question of uniqueness (up to isomorphism) of the representation
$ \pi' $.
\par\bigskip\noindent
{\bf Proposition 3.} {\it Let $ ({\cal H}',\pi',\Omega') $ and $ ({\cal H}'',
\pi'',\Omega'') $ be two graded
representations of $ {\cal A} $ in $\bBZ_2$ graded Hilbert
spaces and let $ \Gamma' $ and $ \Gamma'' $ are the operators represented
both as $\left( \begin{array}{cc} I&0\\ 0&-I\end{array}\right) $ relative to the respective
decompositions of $ {\cal H}' $ and $ {\cal H}'' $. Suppose that the
two representations satisfy $ (\Omega',\Gamma'\pi'(a)\Omega')=(\Omega'',\Gamma''\pi''(a)\Omega'')
 $ and $ (\Omega',\pi'(a)\Omega')=(\Omega'',\pi''(a)\Omega'')\ \forall a\in{\cal A}\ $.
Then $ \pi'$ and $ \pi'' $ are unitarily equivalent as
graded representations with an intertwining map $ V $ that respects the
gradings of the two Hilbert spaces }\p\smallskip\noindent
\section{The Unbounded Case}
As mentioned in the Introduction, there is strong evidence that the framework of bounded graded-KMS functionals may be too restrictive. In particular, it is incompatible with a requirement for locality in infinite-volume quantum field theory \cite{Longo, Buch}. Thus it makes sense to try to generalize the results in the last section to the case of unbounded graded-KMS functionals.\par
We will assume (see, e.g., \cite{Haag}, Ch. III.) that we are given a {\it net of $C^*$-algebras} 
$$
{\cal O}\rightarrow {\cal A(O)}
$$
assigning to each bounded open region in space-time ${\cal O}$ a $C^*$-algebra ${\cal A(O)}$ with the property
$$
{\cal O}_1\subset {\cal O}_2\ \ \Rightarrow \ \ {\cal A(O}_1)\subset {\cal A(O}_2)\ .
$$
The {\it algebra of local ``observables''} is defined as
$${\cal A}_{\hbox{\tiny loc}}:=\bigcup_{\cal O} {\cal A(O)}$$
and the algebra ${\cal A}$ is the completion of ${\cal A}_{\hbox{\tiny loc}}$.\par\noindent
{\bf Note} Strictly speaking, from the point of view of Algebraic Quantum Field Theory, odd operators with respect to the decomposition of the Hilbert space into bosonic and fermionic sectors are not observables.\par
We assume that (${\cal A}$, $\alpha_t$, $\gamma$) is a $\bBZ_2$ graded $C^*$-dynamical system with the grading $*$-automorphism $\gamma$ preserving each local algebra ${\cal A(O)}$ and commuting with the one-parameter $*$-automorphism group $\alpha_t$.
Following (almost exactly) \cite{Buch}, we adopt the following
\par \noindent
{\bf Definition:} The functional $\omega$ will be called {\it unbounded graded-KMS functional} whenever its domain 
$\hbox{Dom}\ \omega$ is a dense $*$-subalgebra of ${\cal A}$, which is $\gamma$ and $\alpha_t$ invariant and the following conditions are satisfied:\par
(A) For any two elements $a,b\in{\cal A}$, such that $a\,\alpha_t(b)\in \hbox{Dom}\ \omega$ and 
$\alpha_t(b)\,a\in \hbox{Dom}\ \omega$ for all $t$, there exists a (unique) complex function $F_{a,b}(z)$ defined on the strip 
$\{z\,|\,0\le\hbox{Im} z\le 1\}$ which is analytic in the interior of that strip and satisfies on the boundaries
\begin{eqnarray}
F_{a,b}(t)=\omega(a\,\alpha_t(b))\label{unb1}\\
F_{a,b}(t+i)=\omega(\alpha_t(b)\,a^{\gamma}) \label{unb2}
\end{eqnarray}\par
(B) For $a,b$ as above we have the following growth condition:
\begin{equation}
|F_{a,b}(t+is)|\le C(1+|t|)^N,\ \ 0\le s\le 1,\label{unb3}
\end{equation}
where $C\in\bBR_+$ and $N\in\bBN$ are constants, depending on $a$ and $b$.\\
{\bf Note} In the case of bounded functionals it is known that  condition (A) above, together with a requirement that $F_{a,b}$ is bounded on the strip is equivalent to the condition in Equation \ref{skms}, which we adopted initially as a definition for the graded-KMS property (see, e.g., \cite{Ped}, 8.12.3 for a proof in the nongraded case). The reason for preferring the current definition for the unbounded case is
that there is no reason to believe that all analytic elements in ${\cal A}$ (in some natural physical context) are in the domain of $\omega$.\par
We will assume that ${\cal A}_{\hbox{\tiny loc}}=\hbox{Dom}\ \omega$. We do not assume that
${\cal A(O)}$ are  unital or that the unit of  ${\cal A}$ is in the domain of $\omega$.  \par
It follows from (A) and (B) that any unbounded graded-KMS functional is $\alpha_t$-invariant and even, i.e., $\gamma$-invariant(\cite{Buch}, Proposition 5.3).\par
In the following considerations we take $\omega$ to be self-adjoint.
Given an unbounded self-adjoint functional $\omega$, one can define a unique unbounded positive functional 
$|\omega |$  by requiring the restriction of $|\omega|$ to any ${\cal A(O)}$ to be the unique modulus of the restriction of 
$\omega$ to that subalgebra. A simple argument, using the structure of the net of local algebras, shows that the definition is unambiguous. Similarly, the Jordan decomposition $\omega=\omega_+-\omega_-$ can be defined in this setting. \par
The automorphism group $\alpha_t$ does not (in general) preserve the local subalgebras ${\cal A(O)}$ separately. In fact it usually has the meaning of time translations acting on the local observables and satisfies $\alpha_t({\cal A(O)})={\cal A(O}^{\,t})$, where 
${O}^{\,t}$ is the time translate of the region ${\cal O}$. It does, however, preserve the whole net of local algebras. From this, the uniqueness of the Jordan decomposition and the $\alpha_t$-invariance of $\omega$ it follows that $\omega_{\pm}$ and
$|\omega|$ are $\alpha_t$-invariant. The grading automorphism $\gamma$ is an automorphism of every ${\cal A(O)}$ separately, so a simpler argument shows that $\omega_{\pm}$ and $|\omega|$ are even.\par
Let $L_{|\omega |}:=\{a\in {\cal A}_{\hbox{\tiny loc}}\ |  \ |\omega |(a^*a)=0\} $, which is a left ideal in ${\cal A}_{\hbox{\tiny loc}}$.The positive functional $|\omega |$ determines an inner product on the space 
${\cal A}_{\hbox{\tiny loc}} /L_{|\omega |}$ and we define the Hilbert space ${\cal H}$ to be the completion of that space.
The algebra ${\cal A}$ is then implemented as an algebra of bounded operators on that Hilbert space (via left multiplication).We use $\pi$ to denote this representation. Let $\eta : {\cal A}_{\hbox{\tiny loc}}\rightarrow {\cal H}$ denote the canonical map, projecting an element of ${\cal A}_{\hbox{\tiny loc}}$ onto ${\cal A}_{\hbox{\tiny loc}} /L_{|\omega |}$, followed by embedding into ${\cal H}$.
It is important to notice that we implement the whole algebra ${\cal A}$ in this way, not just ${\cal A}_{\hbox{\tiny loc}}$. Although for $a\in{\cal A}$ and $b\in{\cal A}_{\hbox{\tiny loc}}$, the product $ab$ is generally not in ${\cal A}_{\hbox{\tiny loc}}$, we can still make sense of the element $\pi(a)\,\eta (b)\in {\cal H}$ since 
$$
||\pi(a)\,\eta (b)||^2=|\omega|(b^*a^*a\,b)\le ||a||^2|\omega|(b^*b)=||a||^2\,||\eta(b)||^2<\infty\ .$$
Strictly speaking the equation above can be given sense by approximating $a$ with elements from ${\cal A}_{\hbox{\tiny loc}}$ and passing to the limit.
\par
Each algebra ${\cal A(O)}$  possesses an approximate unit $u_{\lambda}$ and one can take (see \cite{Ped}, Section 3.3) the element $\Omega_{\cal O}:=\lim_{\lambda} \eta(u_{\lambda})$. We get a net of Hilbert subspaces ${\cal H(O)}:=(\pi({\cal A(O)})\,\Omega_{\cal O})^-\subset {\cal H}$. The restriction of  $\pi({\cal A(O)})$ to 
${\cal H(O)}$ is a standard GNS representation with a cyclic and separating vector $\Omega_{\cal O}$. In particular for any 
$a\in {\cal A(O)}$ we have
$$
|\omega|(a)=(\Omega_{\cal O},\pi(a)\,\Omega_{\cal O})\ 
$$
and there are projections $p_{{\cal O}\pm}\in( \pi({\cal A}))'$ as in Section 2, relating (the restrictions of ) $\omega_{\pm}$ and 
$|\omega|$. It is easy to see that the nets $p_{{\cal O}+}$ and $p_{{\cal O}-}$ are nondecreasing and therefore there are limits $p_+$ and $p_-$ in $(\pi({\cal A}))'$ so that for any $a\in {\cal A(O)}$ we will have Equations \ref{omegapm} and \ref{Gamma} with $\Omega$ replaced by $\Omega_{\cal O}$. 
As in Section 2, we take the weak closure ${\cal B}:=\pi({\cal A})''$ and note that ${\cal B}$ inherits a net of local von Neumann subalgebras ${\cal B(O)}$. The functionals $|\omega|$, $\omega$ and $\omega_{\pm}$ are extended to ${\cal B}_{\hbox{\tiny loc}}:=\bigcup {\cal B(O)}$ in the obvious way, applying the analogs of Equation \ref{omegas}. Using the results for normal functionals on the local algebras ${\cal B(O)}$ we get nets of projections $\chi_{{\cal O}+}$ and $\chi_{{\cal O}-}$ and their limits $\chi_+$ and $\chi_-$, so that Equations \ref{Sakai1} and \ref{Sakai2} remain valid in this more general context. It is quite clear that
$\chi_{{\cal O}+}+\chi_{{\cal O}-}=1_{\cal B(O)}$ (the unit of ${\cal B(O)}$) and by passing to the limit we have 
$\chi_+ + \chi_- =1$ and similarly $p_+ + p_- = 1$.\par
The projections $\chi_{\pm}$ and $p_{\pm}$ must be even and $\alpha_t$-invariant. The former is obvious. The latter is due to the invariance of $\omega$ and $\omega_{\pm}$ and the uniqueness of Sakai's polar decomposition on each ${\cal B(O)}$. We should emphasize that each
$\chi_{{\cal O}+}$ and $\chi_{{\cal O}-}$ is not $\alpha_t$-invariant, but $\alpha_t$ induces automorphisms of the whole nets
of projections and the limits $\chi_{\pm}$ are invariant.
\par\bigskip\noindent
{\bf Proposition 4.} {\it Let $\omega$ be a self-adjoint unbounded graded-KMS functional in the context of the preceding paragraphs and let $|\omega|$ and $\omega_{\pm}$ be the modulus and the positive and negative parts of $\omega$, respectively, in that context. Then $|\omega|$ is an unbounded KMS functional with the same domain. (The definition of the latter should be clear --- it is the same as the definition for a graded-KMS functionals, without the grading automorphism $\gamma$ in} Equations \ref{unb1} and \ref{unb2}.) {\it If $ a\in{\cal A}_+ $ (i.e., $a$ is an even element) and $a$ is in the left kernel of 
$ \omega_+ $, then $ a^* $ is in the left kernel of $ \omega_+ $ as well.
The same is true for $ \omega_- $. If $ a\in{\cal A}_- $ (i.e., $a$ is an odd element) and $a$ is in the left kernel of
$ \omega_+ $, then $ a^* $ is in the left kernel of $ \omega_- $ and vice
versa.
}\par\smallskip\noindent
{\bf Proof:} The proof relies on identities, analogous to Equations \ref{id1} and \ref{id2}. To avoid unnecessary complications we will
only consider elements $a,b\in{\cal A}$. The projections $\chi_{\pm}$ are not in ${\cal A}$ but in ${\cal B}$, so Equations 
\ref{id1} and \ref{id2} make sense in the GNS-representation described in this section. 
First we will need to extend $\omega$ to a certain class of elements outside of  ${\cal A}_{\hbox{\tiny loc}}$, which will be analytic with respect to $\alpha_t$. Recall that an element $a$ is called analytic if $\alpha_t(a)$ has an extension to an entire function $\alpha_z(a)$.
For every $a\in{\cal A}_{\hbox{\tiny loc}}$ define a family of elements (see \cite{Ped}, 8.12.1)
\begin{equation}
a_{\sigma ,z}:=(\pi)^{-1/2}\,\sigma^{-1}\int \alpha_t(a)\,\exp (-(t-z)^2/\sigma^2)\,dt\ ,\quad\sigma\in\bBR,\quad z\in\bBC\  .
\label{analytic}
\end{equation}
We have $a_{\sigma ,0}\to a$ as $\sigma\to 0$ (norm convergence) and $\alpha_z(a_{\sigma ,\zeta})=a_{\sigma ,z+\zeta}$.
Notice that the elements $a_{\sigma ,z}$ are generally not in ${\cal A}_{\hbox{\tiny loc}}$ but can of course be approximated by elements from ${\cal A}_{\hbox{\tiny loc}}$, for example by taking Riemann sums over increasing intervals in place of the integral
in Equation \ref{analytic}.
\par
We take $\tilde{\cal A}_{\hbox{\tiny loc}}$ to be the algebra generated by ${\cal A}_{\hbox{\tiny loc}}$ and elements of the type
$a_{\sigma,z}$ (in the sense of sums of finite products). The functionals $\omega$, $|\omega|$ and $\omega_{\pm}$ are extended in an obvious way to $\tilde{\cal A}_{\hbox{\tiny loc}}$. For example for $a,b\in{\cal A}_{\hbox{\tiny loc}}$ and $c_{\sigma,z}$, $d_{\sigma ',z'}$ as above, we define
$$
\omega(a\,c_{\sigma,z}b\,d_{\sigma ',z'}):=(\pi\sigma\sigma')^{-1}
\int\!\!\int dt\,dt' \exp(-(t-z)^2/\sigma^2)\,\exp(-(t'-z')^2/\sigma'^2)\,\omega(a\,\alpha_t(c)\,b\,\alpha_{t'}(d))\ ,
$$
which makes sense due to the growth condition Equation \ref{unb3}. Furthermore, for every $a_{\sigma, z}\in \tilde{\cal A}_{\hbox{\tiny loc}}$, taking a sequence $a_n\to a_{\sigma, z},\ a_n\in{\cal A}_{\hbox{\tiny loc}}$, we have
$$
\omega_{\pm}(a_n^*a_n)\ge 0\ \ \Rightarrow \lim_{n\to\infty }\omega_{\pm}(a_n^*a_n)=\omega_{\pm}(a_{\sigma,z}^* a_{\sigma,z})\ \ge 0\ ,
$$
which shows that $\omega_{\pm}$ and similarly $|\omega|$ remain positive on $\tilde{\cal A}_{\hbox{\tiny loc}}$.
\par
The functional $\omega$ is graded-KMS on $\tilde{\cal A}_{\hbox{\tiny loc}}$. Indeed, for $a\in{\cal A}_{\hbox{\tiny loc}}$ and $b'=b_{\sigma,\zeta}$ the function 
\begin{eqnarray}
F_{a,b'}(t):=\omega(a\,\alpha_t(b'))
=\int \exp(-(t'-\zeta)^2/\sigma^2)\,\omega(a\,\alpha_{t+t'}(b))\,dt'\nonumber\\
=\int \exp(-(t'-\zeta)^2/\sigma^2)\,F_{a,b}(t+t')\,dt'\nonumber
\end{eqnarray}
extends to the entire function 
$F_{a,b'}(z)=\omega(a\,\alpha_z(b'))=\omega(a\,b_{\sigma,z+\zeta})$ and 
\begin{eqnarray}
F_{a,b'}(t+i)=\omega(a\,\alpha_{t+i}(b'))=\int \exp(-(t'-\zeta)^2/\sigma^2)\,F_{a,b}(t+t'+i)\,dt'\nonumber \\
=\int \exp(-(t'-\zeta)^2/\sigma^2)\,\omega(\alpha_{t+t'}(b)\,a^{\gamma})\,dt'=\omega(\alpha_t(b')\,a^{\gamma})
\ ,\nonumber
\end{eqnarray}
where we used the graded-KMS condition Equations \ref{unb1} and \ref{unb2} on  ${\cal A}_{\hbox{\tiny loc}}$ in the last line.
\par
The analogs of  Equations \ref{id1} and \ref{id2} for arbitrary $a,b\in\tilde{\cal A}_{\hbox{\tiny loc}}$ are proven as follows.
Take, e.g., $a$ to be odd analytic in $\tilde{\cal A}_{\hbox{\tiny loc}}$ and consider for example
\begin{eqnarray}
0\le\omega_+(a^*\chi_+\,a)=\omega_+(a^*\chi_+\,a\,\chi_+)
=\omega(a^*\chi_+\,a\,\chi_+)
=\omega(\alpha_{-i}(a\,\chi_+)(a^*\chi_+)^{\gamma})\nonumber\\
=-\omega(\alpha_{-i/2}(a)\,\chi_+\alpha_{i/2}(a^*)\chi_+)
=-\omega_+(\alpha_{-i/2}(a)\,\chi_+(\alpha_{-i/2}(a))^*)\le 0\,.\nonumber
\end{eqnarray}
This inequality shows that $\omega_+(a^*\chi_+\,a)=0$.
We have used the fact that both $a^*\chi_+\,a$ and $\alpha_{-i/2}(a)\,\chi_+(\alpha_{-i/2}(a))^*$ are positive
elements of $\tilde{\cal A}_{\hbox{\tiny loc}}$, the $\alpha_t$-invariance of $\omega$ and $\chi_+$ as well as $\chi_+$ being odd and the property of $*$-conjugation $(\alpha_z(a))^*=\alpha_{\bar{z}}(a^*)$.
Now, for arbitrary elements $a,b\in{\cal A}_{\hbox{\tiny loc}}$, $a$ -- odd, we show first that $\omega_+(a^*\chi_+\,a)=0$ by approximating $a$ with analytic elements from $\tilde{\cal A}_{\hbox{\tiny loc}}$ and then, using Schwarz inequality, we get
$\omega_+(b\chi_+\,a)=0$
\par
Showing that $|\omega|$ is a KMS functional is already easy. Take, e.g., $a$ -- odd and consider the function
$G_{b,a}(t):=|\omega|(b\,\alpha_t(a))$. We need only look at the case $b$ -- odd, since $|\omega|$ is even functional. We calculate
$$
\begin{array}{rl}
|\omega|(b\,\alpha_t(a))&=\omega_+(b\,\chi_+\,\alpha_t(a))+\omega_+(b\,\chi_-\,\alpha_t(a))
+\omega_-(b\,\chi_+\,\alpha_t(a))+\omega_-(b\,\chi_-\,\alpha_t(a))\nonumber \\
&=\omega_+(b\,\chi_-\,\alpha_t(a))-\omega_-(b\,\chi_-\,\alpha_t(a))
-\omega_+(b\,\chi_+\,\alpha_t(a))+\omega_-(b\,\chi_+\,\alpha_t(a))\nonumber \\
&=-\omega(b\,(\chi_+-\chi_-)\,\alpha_t(a))=-F_{b(\chi_+-\chi_-),a}(t)\ .\nonumber
\end{array}\nonumber
$$
The first and the fourth terms in the right-hand side of the first line are zero, so we have switched the signs in front of them in the next line. We know that there is an analytic extension of $F_{b(\chi_+-\chi_-),a}(t)$ to the strip $\{z\,|\,0\le\hbox{Im} z\le 1\}$ and thus the same is true for $G_{b,a}(t)$ and 
\begin{equation}
G_{b,a}(t+i)=-F_{b(\chi_+-\chi_-),a}(t+i)=-\omega(\alpha_t(a)\,b^{\gamma}(\chi_+-\chi_-))=|\omega|(\alpha_t(a)\,b)\ .
\end{equation}
A similar argument demonstrates the KMS property for $|\omega|$ when $a$ and $b$ are even.\par
The last statement of the proposition is a simple consequence of the analogs of Equations  \ref{id1} and \ref{id2}. For example, suppose $a$ is odd and $a$ is in the left kernel of $\omega_+$. This means that $a=a\,\chi_-$ and therefore 
$$
\omega_-(a\,a^*)=\omega_-(a\,\chi_-\,a^*)=0\ .
$$
The proof is complete.
\par\smallskip
Finally, the structure can be completed by defining the modular conjugation operator $J$. For this we first extend by continuity the canonical map $\eta:{\cal A}_{\hbox{\tiny loc}}\to{\cal H}$ to $\tilde{\cal A}_{\hbox{\tiny loc}}$. This is possible, since for any 
$a\in\tilde{\cal A}_{\hbox{\tiny loc}}$ we saw that $|\omega|(a^*a)<\infty$. For any analytic element $a\in\tilde{\cal A}_{\hbox{\tiny loc}}$ we define
$$
J\,\eta(a):=\eta(\alpha_{\frac i2}(a^*))\ .
$$
This defines $J$ on a dense subset of ${\cal H}$. The proof that $J$ is antiunitary is the same as before. Taking an operator of complex conjugation $K$ and defining the unitary map $U:=KJ$, we define a representation $\pi':=U\pi U^*$.\par
The orthogonal decomposition ${\cal H}={\cal H}_+ \oplus{\cal H}_-$ with ${\cal H}_{\pm}:=p_{\pm}{\cal H}$ is compatible with the unbounded self-adjoint functional $\omega$. Equations \ref{maps} remain valid and it becomes clear that
the main result of Section 3 (Proposition 2) now extends to the case of $\omega$ being an unbounded graded-KMS functional.

\end{document}